\documentclass[12pt,preprint]{aastex}

\slugcomment{draft of \today}
\shorttitle{IGM and Arrival Direction of UHE Protons}
\shortauthors{Ryu et al.}

\font\mbf = cmmib10 scaled\magstep1
       \font\mbfs = cmmib10 \font\mbfss = cmmib10 scaled 833
       \def\bmit{\fam9 }
\font\msybf = cmbsy10 scaled\magstep1
       \font\msybfs = cmbsy10 \font\msybfss = cmbsy10 scaled 833
       \def\bmsy{\fam10 }

\begin{document}

\title{Intergalactic Magnetic Field and
       Arrival Direction of Ultra-High-Energy Protons}

\author{Dongsu Ryu\altaffilmark{1,4}, Santabrata Das\altaffilmark{2},
        and Hyesung Kang\altaffilmark{3}}
\altaffiltext{1}{Department of Astronomy and Space Science,
                 Chungnam National University, Daejeon 305-764, Korea;
                 \\ ryu@canopus.cnu.ac.kr}
\altaffiltext{2}{Department of Physics, Indian Institute of Technology
                 Guwahati, Guwahati, 781039, Assam, India;
                 \\ sbdas@iitg.ernet.in}
\altaffiltext{3}{Department of Earth Sciences, Pusan National University,
                 Pusan 609-735, Korea; \\ kang@uju.es.pusan.ac.kr}
\altaffiltext{4}{Corresponding Author}

\begin{abstract}

We studied how the intergalactic magnetic field (IGMF) affects the
propagation of super-GZK protons that originate from extragalactic
sources within the local GZK sphere. 
Toward this end, we set up hypothetical sources of ultra-high-energy
cosmic-rays (UHECRs), virtual observers, and the magnetized cosmic web
in a model universe constructed from cosmological structure formation
simulations.
We then arranged a set of reference objects mimicking active galactic
nuclei (AGNs) in the local universe, with which correlations of 
simulated UHECR events are analyzed. 
With our model IGMF, the deflection angle between the arrival direction
of super-GZK protons and the sky position of their actual
sources is quite large with the mean value of
$\langle \theta \rangle \sim 15^{\circ}$ and the median value of
$\tilde \theta \sim 7 - 10^{\circ}$.
On the other hand, the separation angle between the arrival direction
and the sky position of nearest reference objects is substantially smaller
with $\langle S \rangle \sim 3.5 - 4^{\circ}$,
which is similar to the mean angular distance in the sky to nearest
neighbors among the reference objects. 
This is a direct consequence of our model that the sources, observers, 
reference objects, and the IGMF all trace the matter distribution 
of the universe.
The result implies that extragalactic objects lying closest to the
arrival direction of UHECRs are not necessary their actual sources. 
With our model for the distribution of reference objects, the fraction of 
super-GZK proton events, whose closest AGNs are true sources, is less
than 1/3.
We discussed implications of our findings for correlation studies
of real UHECR events.

\end{abstract}


\section{Introduction}

The nature and origin of ultra-high-energy cosmic rays (UHECRs), 
especially above the so-called Greisen-Zatsepin-Kuz'min (GZK) energy of 
$E_{\rm GZK}\approx$ 50 EeV (1 EeV $=10^{18}$ eV), has been one of
most perplexing puzzles in astrophysics over five decades 
and still remains to be understood \citep[see][for a review]{nw00}.
The highest energy CR detected so far is the Fly's Eye event with
an estimated energy of $\sim 300$ EeV \citep{flyseye94}.
At these high energies protons and nuclei cannot be confined 
and accelerated effectively within our Galaxy, 
so the sources of UHECRs are likely to be extragalactic.

At energies higher than $E_{\rm GZK}$, it is expected that protons lose
energy and nuclei are photo-disintegrated via the interactions with
the cosmic microwave background radiation (CMB) along their trajectories
in the intergalactic space \citep{grei66,zk66,psb76}.
The former is known as the GZK effect.
So a significant suppression in the energy spectrum above $E_{\rm GZK}$
could be regarded as an observational evidence
for the extragalactic origin of UHECRs \citep[see, e.g.,][]{bgg06}.
However, the accurate measurement of the UHECR spectrum is very difficult, 
partly because of extremely low flux of UHECRs.
But a more serious hurdle is the uncertainties in the energy calibration 
inherent in detecting and modeling extensive airshower events  
\citep[e.g.,][]{nw00, watson06}.
Nevertheless, both the Yakutsk Extensive Air Shower Array (Yakutsk)
and the High Resolution Fly's Eyes (HiRes) reported observations of
the GZK suppression \citep{yakutsk04,hires08a}, while the Akeno Giant
Air Shower Array (AGASA) claimed a conflicting finding of no suppression
\citep{agasa04}.
More recent data from the Pierre Auger Observatory (Auger) support
the existence of the GZK suppression \citep{auger08b,auger09c}.
Below $E_{\rm GZK}$, however, the four experiments reported the fluxes 
that are different from each other by up to a factor of several, implying
the possible existence of systematic errors in their energy calibrations
\citep{bere09}.

The overall sky distribution of the arrival directions of UHECRs 
below $E_{GZK}$ seems to support the isotropy hypothesis 
\citep[see, e.g.,][]{nw00}.
This is consistent with the expectation of uniform distribution of
extragalactic sources;
the interaction length (i.e. horizon distance) of protons below
$E_{GZK}$ is a few Gpc and the universe can be considered homogeneous
and isotropic on such a large scale. 
The horizon distance for super-GZK events, however, decreases sharply
with energy and $R_{\rm GZK} \sim 100 $ Mpc for $E=100$EeV \citep{bg88}. 
The matter distribution inside the local GZK horizon ($R_{\rm GZK}$) is
inhomogeneous.
Since powerful astronomical objects are likely to form at deep
gravitational potential wells, we expect the distribution of the UHECR
sources would be inhomogeneous as well.
Hence, if super-GZK proton events point their sources, their arrival
directions should be anisotropic. 

The anisotropy of super-GZK events, hence, has been regarded to provide
an important clue that unveils the sources of UHECRs.
So far, however, the claims derived from analyses of different experiments
are often tantalizing and sometimes conflicting. 
For instance, with an excessive number of pairs and one triplet
in the arrival direction of CRs above 40 EeV, the AGASA data
support the existence of small scale clustering \citep{agasa96,agasa99}.
On the other hand, the HiRes stereo data are consistent with the hypothesis
of null clustering \citep{hires04,hires09}.
The auto-correlation analysis of the Auger data reported a weak excess
of pairs for $E > 57$ EeV \citep{auger08a}.
In addition, the Auger Collaboration found a correlation
between highest energy events and the large scale structure (LSS) of the
universe using nearby active galactic nuclei (AGNs) in the \citet{vcv06}
catalog \citep{auger07a,auger08a,auger09b} as well as using nearby
objects in different catalogs \citep{auger09a}.
A correlation between highest AGASA events with nearby galaxies from SDSS
was reported \citep{tns09}.
The HiRes data, however, do not show such correlation of highest energy
events with nearby AGNs \citep{hires08b}, but instead show a correlation
with distant BL Lac objects \citep{hires06}.

The interpretation of anisotropy and correlation analyses is, however,
complicated owing to the intervening galactic magnetic field (GMF) and
intergalactic magnetic field (IGMF);
the trajectories of UHECRs are deflected
by the magnetic fields as they propagate through the space between
sources and us, and hence, their arrival directions are altered.
Even with considerable observational and theoretical efforts, however,
the nature of the GMF and the IGMF is still poorly constrained.
Yet, models for the GMF generally assume a strength of $\sim$ a few $\mu$G
and a coherence length of $\sim 1$ kpc for the field in the Galactic halo
\citep[see, e.g.,][]{stan97}, and predict the deflection of UHE protons
due to the GMF to be $\theta \sim$ a few degrees \citep[see, e.g.,][]{ts08}.
The situation for the IGMF in the LSS has been confusing.
Adopting a model for the IGMF with the average strength of
$\langle B \rangle \sim 100$ nG in filaments, \citet{sme03} showed that
the deflection of UHECRs due to the IGMF could be very large, e.g.,
$\theta > 20^{\circ}$ for protons above 100 EeV.
On the other hand, \citet{dgst05} adopted a model with
$\langle B \rangle \sim 0.1$ nG in filaments and showed that the
deflection should be negligible, e.g., $\theta \ll 1^{\circ}$ for protons
with 100 EeV.

Recently, \citet{rkcd08} proposed a physically motivated model for the
IGMF, in which a part of the gravitational energy released during
structure formation is transferred to the magnetic field energy as a
result of the turbulent dynamo amplification of weak seed fields
in the LSS of the universe.
In the model, the IGMF follows largely the matter distribution in the
cosmic web, and the strength and coherence length are predicted to be
$\langle B \rangle \sim 10$ nG and $\sim 1$ Mpc for the field in filaments.
Such field in filaments is expected to induce the Faraday rotation
\citep{cr09}, which is consistent with observation \citep{xkhd06}.
With this model IGMF, \citet{dkrc08} (Paper I hereafter) calculated 
the trajectories of UHE protons ($E>10$ EeV) that were injected at
extragalactic sources associated with the LSS in a simulated model universe.  
We then estimated that only $\sim 35$ \% of UHE protons above 60 EeV
would arrive at us with $\theta \leq 5^{\circ}$ and the average value of
deflection angle would be $\langle \theta \rangle \sim 15^{\circ}$.
Note that the deflection angle of $\langle \theta \rangle \sim 15^{\circ}$
is much larger than the angular window of $3.1^{\circ}$ used by the Auger
collaboration in the study of the correlation between highest energy
UHECR events and nearby AGNs \citep{auger07a,auger08a,auger09b}.

In this contribution, as a follow-up work of Paper I, we investigate
the effects of the IGMF on the arrival direction of super-GZK protons
above 60 EeV coming from sources within 75 Mpc.
The limiting parameters for energy and source distance are chosen to
match the recent analysis of the Auger collaboration. 
Without knowing the true sources of UHECRs, the statistics that can
be obtained with observational data from experiments are limited;
some statistics that are essential to reveal the nature of sources
are difficult or even impossible to be constructed.
On the other hand, with data from simulations, any statistics can be
explored.
In that sense, simulations complement experiments.
Here, with the IGMF suggested by \citet{rkcd08}, we argue that the large
deflection angle of super-GZK protons due to the IGMF is not inconsistent
with the anisotropy and correlation recently reported by the Auger
collaboration.
However, the large deflection angle implies that the nearest object to a
UHECR event in the sky is not necessarily its actual source.
In Section 2, we describe our models for the LSS of the universe, IGMF,
observers, and sources of UHECRs, reference objects for correlation
study, and simulations.
In Section 3, we present the results, followed by
a summary and discussion in Section 4.

\section{Models and Simulations}

In our study, the following elements are necessary: 
1) a model for the IGMF on the LSS,
2) a set of virtual observers that represent ``us'', an observer at the
Earth, in a statistical way, 
3) a set of hypothetical sources of UHE protons with a specified injection
spectrum, and
4) a set of reference objects with which we performed a correlation study
of simulated events.
In Paper I, we described in detail how we set up 1), 2), and 3) by using
data from cosmological structure formation simulations.
Below, we briefly summarize models for 1), 2), and 3) and explain in
details the reason to introduce ``reference objects'' in this study.  

\subsection{Large Scale Structure of the Universe}

We assumed a concordance $\Lambda$CDM model with the following parameters: 
$\Omega_{\rm BM}=0.043$, $\Omega_{\rm DM}=0.227$, and
$\Omega_{\Lambda}=0.73$, $h \equiv H_0$/(100 km/s/Mpc) = 0.7,
and $\sigma_8 = 0.8$.
The model universe for the LSS was generated through simulations in a cubic
region of comoving size $100 h^{-1} (\equiv 143)$ Mpc with $512^3$ grid
zones for gas and gravity and $256^3$ particles for dark matter, using a
PM/Eulerian hydrodynamic cosmology code described in \citet{rokc93}.
The simulations have a uniform spatial resolution of $195.3 h^{-1}$ kpc.
The standard set of gasdynamic variables, the gas density, $\rho_g$,
temperature, $T$, and the flow velocity, ${\bmit v}$, were used to calculate
the quantities required in our model such as the X-ray emission weighted
temperature $T_X$, the vorticity, ${\bmsy \omega}$, and the turbulent
energy density, $\varepsilon_{\rm turb}$, at each grid. 

\subsection{Intergalactic Magnetic Field}

We adopted the IGMF from the model by \citet{rkcd08};
the model proposes that turbulent-flow motions are induced via the cascade
of the vorticity generated at cosmological shocks during the formation of
the LSS of the universe, and the IGMF is produced as a consequence of the
amplification of weak seed fields of any origin by the turbulence.
Then, the energy density (or the strength) of the IGMF can be estimated
with the eddy turnover number and the turbulent energy density as follow:
\begin{equation}
\varepsilon_B = \phi \left({t \over t_{\rm eddy}}\right)
\varepsilon_{\rm turb}.
\end{equation}
Here, the eddy turnover time is defined as the reciprocal of the vorticity
at driving scales, $t_{\rm eddy} \equiv 1/\omega_{\rm driving}$ 
(${\bmsy \omega} \equiv {\bmsy \nabla}\times{\bmit v}$), and $\phi$ is the
conversion factor from turbulent to magnetic energy that depends on the
eddy turnover number $t/t_{\rm eddy}$.
The eddy turnover number was estimated as the age of universe times the
magnitude of the local vorticity, that is, $t_{\rm age}\ \omega$.
The local vorticity and turbulent energy density were calculated
from cosmological simulations for structure formation described above.
A functional form for the conversion factor was derived from a separate,
incompressible, magnetohydrodynamic (MHD) simulation of turbulence dynamo.

For the direction of the IGMF, we used that of the passive fields from
cosmological simulations, in which magnetic fields were generated through
the Biermann battery mechanism \citep{biermann50} at cosmological shocks
and evolved passively along with flow motions \citep{kcor97,rkb98}.
In principle, if we had performed full MHD simulations, we could have
followed the amplification of the IGMF through turbulence dynamo.
In practice, however, the currently available computational resources
do not allow a numerical resolution high enough to reproduce the full 
development of MHD turbulence.
Since the numerical resistivity is larger than the physical resistivity
by many orders of magnitude, the growth of magnetic fields is saturated
before dynamo action becomes fully operative \citep[see, e.g.,][]{kcor97}.
This is the reason why we adopted the model of \citet{rkcd08} to estimate 
the strength of the IGMF, but we still used the the passive fields from
cosmological simulations to model the field direction.

Figure 1 shows the distribution of magnetic field strength in a slice of
(143 Mpc)$^2$ in our model universe.
It shows that the IGMF is structured like the matter in the cosmic web.
As a matter of fact, the distribution of the IGMF is very well correlated
with that of matter.
The strongest magnetic field of $B \ga 0.1\mu$G is found in and
around clusters, while the field is weaker in filaments, sheets, and
voids.
In filaments which are mostly composed of the warm-hot intergalactic
medium (WHIM) with $T = 10^5 - 10^7$ K, the IGMF has
$\langle B\rangle \sim 10$ nG and $\langle B^2\rangle^{1/2} \sim$
a few $\times~10$ nG \citep{rkcd08}.
Note that the deflection of UHECRs arises mostly due to the field in
filaments (see Paper I).
The energy density of the IGMF in filaments is
$\varepsilon_B \sim 10^{-16}$ ergs cm$^{-3}$, which is a few times smaller
than the gas thermal energy density and an order of magnitude smaller
than the gas kinetic energy density there.\footnote
{We note that our model does not include a possible contribution to the
IGMF from galactic black holes, AGN feedback \citep[see, e.g.,][]{kdlc01};
so our model may be regarded to provide a baseline for the IGMF.
With the contribution, the real IGMF might be even stronger, resulting in
even larger deflection (see Section 3.1).}
The IGMF in filaments induces the Faraday rotation; the root-mean-square
(rms) value of rotation measure (RM) is predicted to be a few rad m$^{-1}$
\citep{cr09}.
That is consistent with the values of RM toward the Hercules and
Perseus-Pisces superclusters reported in \citet{xkhd06}.\footnote
{The values of $|$RM$|$ in \citet{xkhd06} is an order of
magnitude larger than the value above, a few rad m$^{-1}$.
However, \citet{xkhd06} quoted the path length, which is about two
orders of magnitude larger.}

\subsection{Observer Locations}

In the study of the arrival direction of UHECRs, the IGMF around us, that is,
in the Local Group, is important too.
It would have been ideal to place ``the observer'' where the IGMF is
similar to that in the Local group.
Unfortunately, however, little is known about the IGMF in the Local Group.
Hence, instead, we placed ``virtual'' observers based on the X-ray emission
weighted temperature $T_X$.
The groups of galaxies that have the halo temperature similar to that of the 
Local Group, 0.05 keV $< kT_X <$ 0.5 keV \citep{rp01}, were identified.
About 1400 observer locations were chosen by the temperature criterion.
In reality, there should be only one observer on the Earth.
But in our modeling we could choose a number of observer locations to
represent statistically ``us'' without loss of generality,  since the
simulated universe is only one statistical representation of the real
universe. 
Then, we modeled observers as a sphere of radius $0.5 h^{-1}$ Mpc located
at the center of host groups, in order to reduce the computing time to
a practical level.
The distribution of handful observers are shown schematically in Figure 1.
One can see that the observers (groups) are not distributed uniformly, 
but instead they are located mostly along filaments.

\subsection{AGNs as Reference Objects}

As noted in Introduction, the Auger Collaboration recently reported
a correlation between the direction of their highest energy events and
the sky position of nearby AGNs from the 12th edition of the \citet{vcv06} 
(VCV) catalog;
the correlation has the maximum significance for UHECRs with $E \ga 60$ EeV
and AGNs with distance $D \la 75$ Mpc \citep{auger07a,auger08a,auger09b}.
There are about 450 AGNs with $D \la 75$ Mpc (more precisely, 442 AGNs
with redshift $z \le 0.018$, for which the maximum significance of the
correlation was found) in the VCV catalog.
In that study it is not known which subclass of those AGNs or what
fraction of them are really true sources of UHECRS.
Here, we regard those AGNs as ``reference objects'', with which correlation
studies are performed. 

In order to compare our correlation study with that of
the Auger collaboration, we specified the following condition
to determine ``model'' reference objects in the simulated universe:
1) the number of the objects within 75 Mpc from each observer should be
on average $\sim 500$,
and 2) their spatial distribution should trace the LSS in a way similar to 
the AGN distribution in the real universe.  
To set up the location of such reference objects, we identified ``clusters''
of galaxies with $kT_X \ga 0.1$ keV in the simulated universe.
Of course some of these clusters with $kT_X \la$ a few keV should be
classified as groups of galaxies. But for simplicity we call all of
them as clusters. 
The reason behind this selection condition is that the gas temperature
is directly  related with the depth of gravitational potential well;
the hottest gas resides in the densest, most nonlinear regions of the LSS
where the most luminous and energetic objects (e.g. AGNs) form through
frequent mergers of galaxies. 
We then assumed that each cluster hosts one reference object at its center.
For each observer, we generated a list of reference objects inside a 
sphere of radius 75 Mpc, whose number is on average $\sim 500$;
the exact number of reference objects varies somewhat for different observers.
Then, each observer has its own sky distribution of reference objects,
with which we studied the correlation of simulated events. 
Although our reference objects could be any astronomical objects that
trace the LSS, hereafter we refer them  as ``model AGNs'', because
the selection criteria were chosen to match the number of AGNs with that
from the VCV catalog.
Figure 1 shows the schematic distribution of handful model AGNs at
the center of host clusters.
Obviously the host clusters (and the AGNs) are not uniformly distributed
either. 

In our set-up, the distance to the model AGNs, $D$, can be arbitrarily small.
In reality, however, the closest AGN in the VCV catalog is NGC 404 at
$D \sim 3$ Mpc in the constellation Andromeda \citep{kkhm04}.
So the model AGNs with the distance from each observer
$D < D_{\rm min} \equiv 3$ Mpc were excluded.

We checked the angular distance $Q$ between a given  reference object to
its nearest neighboring object.
For a set of 442 objects (the number of the AGNs with $z \le 0.018$ in
the VCV catalog), if they are distributed isotropically over the sky of
$4\pi$ radian, the average value of $Q$ would be
$\langle Q_{\rm iso} \rangle \approx 11^{\circ}$.
With the 442 AGNs from the VCV catalog, on the other hand,
$\langle Q_{\rm VCV} \rangle = 3.55^{\circ}$.
The fact that $\langle Q_{\rm VCV} \rangle < \langle Q_{\rm iso} \rangle$
means that the distribution of the AGNs from the VCV catalog is not
isotropic, but highly clustered, following the matter distribution in
the LSS of the universe.
We note that $\langle Q_{\rm VCV} \rangle = 3.55^{\circ}$ is similar to
the angular window of $3.1^{\circ}$ used in the Auger study.
Clearly this agreement is not accidental, but rather consequential.
For the sets of $\sim 500$ model AGNs in our simulations, the average
angular distance is
$\langle Q_{\rm AGN} \rangle = 3.68^{\circ} \pm 1.66^{\circ}$.
The error was estimated with $\langle Q_{\rm AGN} \rangle$ for $\sim 1400$
observers.
That fact that $\langle Q_{\rm AGN} \rangle \sim \langle Q_{\rm VCV} \rangle$
indicates that the spatial clustering of our model AGNs is on average
comparable to that of the AGNs from the VCV catalog.
This provides a justification for our selection criteria for model AGNs
in the simulated universe.
We note that $Q$ is an intrinsic property of the distribution of the
reference objects in the sky and has nothing to do with UHECRs.

\subsection{Sources of UHECRs}

Although AGN is one of viable candidates that would produce UHECRs
\citep[see, e.g.,][for the list of viable candidates]{nw00}, there is
no compelling reason that all the nearby AGNs are the sources of UHECRs.
In this paper, we considered three models with different numbers of sources,
$N_{\rm src}$ (see Table 1), to represent different subsets of AGNs.
1) Among AGNs, radio galaxies are considered to be the most promising
sources of UHECRs \citep[see, e.g.,][]{bs87}, and there are 28 known radio
galaxies within $D \le 75$ Mpc.
So in Model C, we considered on average 28 model AGNs located at 28
hottest host clusters ($kT_x\ga 0.8$keV) within a sphere of radius 75 Mpc
as true sources of UHECRs.
2) Based on the ratio of singlet to doublet events, on the other hand,
\citet{auger08a} argued that the lower limit on the number of sources
of UHECRs would be around 61.
Following this claim, in Model B, we regarded on average 60 model AGNs
located at 60 hottest host clusters ($kT_x\ga 0.55$keV) as true
sources of UHECRs.
2) In Model A, we regarded all the model AGNs (reference objects) as
true sources of UHECRs. 

\begin{deluxetable}{ccccccccc}
\tablecolumns{2}
\tablewidth{0pc}
\tablecaption{Models of different numbers of sources}
\tablehead{
\colhead{Model} & \colhead{$N_{\rm src}$} & \colhead{host clusters} &
\colhead{$\langle \theta \rangle$\tablenotemark{a}} & \colhead{$\tilde \theta$}
& \colhead{$\langle S_{\rm sim} \rangle$\tablenotemark{a}} &
\colhead{$\tilde S_{\rm sim}$ } } 
\startdata
 A & $\sim$500 & $kT_X\ga$ 0.1keV  &13.98  & 7.01 & 3.58 & 2.80 \\
 B & 60 &  $kT_X\ga$ 0.55keV  &15.33  & 8.80 & 3.97 & 3.19\\
 C & 28 &  $kT_X\ga$ 0.8keV  &17.76  & 10.45 & 4.23 & 3.43\\
\enddata
\tablenotetext{a} {Deflection angle $\theta$ and separation angle $S$ are
defined in Section 3}
\end{deluxetable}

\subsection{Simulations of Propagation of UHE Protons}

At sources, UHE protons were injected with power-law energy spectrum;
$N_{\rm inj}(E_{\rm inj})\propto E^{-\gamma}_{\rm inj}$ for
$6 \times 10^{19}$ eV $\le E_{\rm inj} \le 10^{21}$ {\rm eV}, where
$\gamma$ is the injection spectral index.
We considered the two cases of $\gamma = 2.7$ and 2.4.
At each source, protons were randomly distributed over a sphere of
radius  $0.5 h^{-1}$ Mpc, and then launched in random directions.

We then followed the trajectories of UHE protons in our model universe
with the IGMF, by numerically integrating the equations of motion;
\begin{equation}
\frac{d{\bmit r}}{dt} = {\bmit v},~~~~~\frac{d {\bmit p}}{dt}
=e\left( {\bmit v} \times {\bmit B}\right),
\end{equation}
where ${\bmit p}$ is the momentum.
During propagation UHE protons interact with the CMB, and the
dominant processes for energy loss are the pion and pair productions.
The energy loss was treated with the continuous-loss approximation
\citep{bgg06}.
The adiabatic loss due to the cosmic expansion was ignored.

We let UHE protons continue the journey, visiting several observers
during flight, until the energy falls to 60 EeV. 
At observers, the events with $E \ge 60$ EeV were recorded and analyzed.

\section{Results}

\subsection{Deflection Angle}

In Paper I, we considered the deflection angle, $\theta$, between the
arrival direction of UHECR events and the sky position of their sources
(see Figure 2).
Obviously this angle can be calculated only when the true sources are known,
which is not the case in experiments.
In the simulated universe, our model AGNs and observers are located in
strongly magnetized regions.
As illustrated in Figure 1, UHECRs first have to escape from magnetic
halos surrounding sources, then travel through more or less void
regions (path 1) or through filaments (path 2), and finally penetrate
into magnetic halos around observers, to reach observers.
So the degree of deflection depends not only on the magnetic field along
trajectories but also on the fields at host clusters and groups of sources
and observers.
Since the gas temperature, the depth of gravitational potential well, and
the magnetic field energy density are related as 
$kT_X \propto \Phi\ \propto \varepsilon_B$ in our model, hotter clusters
and groups would have stronger fields.
So we expect that if sources and observers are located at hotter hosts,
$\theta$ would be larger on average.

Figure 3 shows $\theta$ versus $D_{\theta}$ for UHE proton events
recorded at observers in our simulations.
Here, $D_{\theta}$ denotes the distance to the actual sources of events. 
The top, middle, and bottom panels are for Models A, B, and C, respectively.
Only the case of injection spectral index $\gamma = 2.7$ is presented.
The case of $\gamma = 2.4$ is similar.
Each dot represents one simulated event, and there are about $10^5$ events
in each Model.
The upper circles connected with dotted lines represent the mean values
of $\theta$ in the distance bins of $[D_\theta, D_\theta+\Delta D_\theta]$.
The mean deflection angle averaged over all the simulated events is
$\langle \theta \rangle = 13.98^{\circ}$,  $15.33^{\circ}$, and
$17.76^{\circ}$ for $\gamma = 2.7$ in Model A, Model B, and Model C,
respectively.
The lower circles connected with solid lines represent the median values
of $\theta$.
The median value for all the simulated events is
$\tilde \theta = 7.01^{\circ}$,  $8.80^{\circ}$, and
$10.45^{\circ}$ for $\gamma = 2.7$ in Model A, Model B, and Model C,
respectively.
The values of $\langle \theta \rangle$ and $\tilde \theta$ for
$\gamma = 2.4$ are similar.
The marks connected with vertical solid lines on the both sides of
median values are the first and third quartiles in the distance bins,
which provide a measure of the dispersion of $\theta$.

We note the following points. 
(1) With our IGMF, the mean deflection angle of UHE protons due to the
IGMF is quite large, much larger than the angular window of $3.1^{\circ}$
used in the Auger correlation study.
It is also much larger than the mean deflection angle that is expected 
to result from the Galactic magnetic field, which is a few degrees
\citep{ts08}.
(2) The mean deflection angle is largest in Model C and smallest in Model A. 
Recall that in Model C sources are located only at 28 hottest clusters,
while in Model A all 500 clusters include sources (see Table 1).
The UHECR events from hotter clusters tend to experience more deflection,
as noted above.
So the mean of deflection angles in the model with hotter host clusters is
larger. 
(3) The mean value of $\theta$ has a minimum at
$D_{\theta,{\rm min}} \sim 20 - 30$ Mpc, which compares to a typical length
of filaments.
As we pointed out in Paper I, this is a consequence of the structured magnetic 
fields that are concentrated along filaments and at clusters. 
In an event with $D_{\theta} < D_{\theta,{\rm min}}$, 
the source and observer are more likely to belong to the same filament,
and so the particle is more likely to travel through strongly magnetized
regions and suffer large deflection (see the path 2 in Figure 1). 
In the opposite regime, the source and observer are likely to belong to
different filaments, so the particle would travel through void regions
(see the path 1 in Figure 1). 
(4) For the events with $D_{\theta} > D_{\theta,{\rm min}}$, the mean and
dispersion of $\theta$ increase with $D_\theta$.
Such trend is expected, since in the diffusive transport model of the
propagation of UHECRs, the deflection angle increases with distance as
$\theta_{\rm rms} \propto \sqrt{D_{\theta}}$
\citep[see, e.g.,][and references therein]{kl09}.
(5) There are more events from nearby sources than from distant sources,
although all the sources inject the same number of UHECRs in our model. 
The smaller number of events for larger $D_\theta$ should be mostly a
consequence of energy loss due to the interaction with the CMB.

\subsection{Separation Angles}

In this study, we also consider the separation angle, $S$, between the
arrival direction of UHECR events and the sky position of nearest reference
objects (see Figure 2).
The angle can be calculated with observation data, once a class of reference
objects (e.g. AGNs, galaxies gamma-ray bursts, and etc.) is specified.
For example, \citet{auger07a} took the AGNs within 75 Mpc in the VCV catalog
as the reference objects for their correlation study. 
However, for a given UHECR event, the nearest AGN in the sky may not be
the actual source; hence, the separation angle between a UHECR event and
its nearest AGN is not necessarily the same as the deflection angle of
the event \citep{hillas09,rkd09}.

We obtained $S$ for simulated events with our model reference objects
(AGNs).
Figure 4 shows $S$ versus $D_S$.
Here, $D_S$ denotes the distance to nearest AGNs.
Again only the case of $\gamma = 2.7$ is presented, and the case of
$\gamma = 2.4$ is similar (see Figure 5).
The circles connected with solid line represent the mean values of $S$ for
the events with nearest AGNs in the distance bins of $[D_S, D_S+\Delta D_S]$.
The mean separation angle averaged over all the simulated events is
$\langle S_{\rm sim} \rangle = 3.58^{\circ}$,  $3.97^{\circ}$, and
$4.23^{\circ}$ for $\gamma = 2.7$ in Models A, B, and C,
respectively.

We note the following points.
(1) The mean separation angle is much smaller than the mean deflection
angle, $\langle S_{\rm sim} \rangle \sim (1/4) \langle \theta \rangle$
(see the next section for further discussion).
(2) The mean separation angle is largest in Model C and smallest in Model A,
although the difference of $\langle S_{\rm sim} \rangle$ among the models
is less than that of $\langle \theta \rangle$.
With larger deflection angles in Model C, there is a higher probability for a
event to be found further away from the region where model AGNs are
clustered, so $S$ is on average larger as well.
(3) Similarly as in the $\theta$ versus $D_{\theta}$ distribution,
the distribution of $\langle S_{\rm sim} \rangle$ has the minimum at
around $D_S \sim 35$ Mpc.
This is again a signature of the filamentary structures of the LSS.
(4) Contrary to $D_{\theta}$, there are more events with larger $D_S$
than with smaller $D_S$.
It is simply because there are more AGNs with larger $D_S$.

\subsection{Comparison with the Auger Data}

We also obtained $S$ for the 27 Auger events of highest energies, published
in \citet{auger08a}, with 442 nearby AGNs from the VC catalog.
In Figure 5, we compare the $S$ versus $D_s$ distribution for the Auger data 
with that of our simulations.
The upper circles connected with dashed/dot-dashed lines represent the
mean values of $S_{\rm sim}$ of simulated events as in Figure 4, but this
time both cases of $\gamma = 2.7$ and 2.4 are presented.
The lower circles connected with solid/dotted lines represent the median
values of $S_{\rm sim}$. 
The difference between the cases of $\gamma = 2.7$ and 2.4 is indeed small.
The median value of $S_{\rm sim}$ for all the simulated events is
$\tilde S_{\rm sim} = 2.80^{\circ}$,  $3.19^{\circ}$, and
$3.43^{\circ}$ for $\gamma = 2.7$ in Models A, B, and C,
respectively.
Asterisks denote the Auger events.
The mean separation angle for the Auger events is
$\langle S_{\rm Auger} \rangle = 3.23^{\circ}$ for 26 events, excluding one
event with large $S$ ($\approx 27^{\circ}$), while
$\langle S_{\rm Auger} \rangle =4.13^{\circ}$  for all the 27 events. 
We note that $\langle S_{\rm Auger} \rangle \sim \langle S_{\rm sim} \rangle$,
even though the mean deflection angle is much larger than the mean separation
angle in our simulations, that is,
$\langle \theta \rangle \gg \langle S_{\rm sim} \rangle$.
In all the models, about a half of the Auger events lie within the quartile
marks: 15, 13, and 13 events for Models A, B, and C, respectively.

With $\langle \theta \rangle \sim 15^{\circ}$ in our simulations, one might
naively expect that such large deflection would erase the anisotropy in
the arrival direction and the correlation between UHECR events and AGNs
(or the LSS of the universe).
However, we argue that the large deflection does not necessarily lead to
the general isotropy of UHECR arrival direction, if the agent of
deflection, the IGMF, traces the local LSS.
Suppose that UHECRs are ejected from sources inside the Local Supercluster.
Some of them will fly along the Supergalactic plane and arrive at the Earth;
their trajectories would be deflected by the magnetic field between sources
and us, but the arrival directions still point toward the Supergalactic plane.
Others may be deflected into void regions, and then they will have less
chance to get reflected back to the direction toward us due to lack of
the turbulent IGMF there.
In a simplified picture, we may regard the irregularities in the IGMF
as the `scatters' of UHECRs;
then the last scattering point will be the arrival direction of UHECRs
\citep[see][for a description of deflection of UHECRs based on
this picture]{kl09}.
As a result, even with large deflection, we still see more UHECRs from
the LSS of clusters, groups, and filaments, and fewer UHECRs from void
regions where both sources and scatters are underpopulated.
Consequently, the anisotropy in the arrival direction of UHECRs can
be maintained and the arrival direction still follows the LSS of the
universe.

Below the GZK energy, the proton horizon reaches out to a few Gpc, 
so the source distribution should look more or less isotropic and the
arrival directions should not show a correlation with nearby AGNs. 
Thus, we do not expect to see anisotropy and correlation for UHECRs
with such energy.

In Section 2.4, we showed that the degree of clustering of our model AGNs
is similar to that of AGNs from the VCV catalog; the mean of the
angular distance $Q$ between a given AGN to its nearest neighboring AGN
is similar, $\langle Q_{AGN} \rangle \sim \langle Q_{\rm VCV}\rangle$. 
In both cases, AGNs follow the matter distribution in the LSS, highly
structured and clustered.
We point that if along with the reference objects, the CR sources and
the IGMF also follow the matter distribution, with
$\langle \theta \rangle \gg \langle S \rangle$,
$\langle S \rangle \sim \langle Q \rangle$ is expected.
The result that 
$\langle S_{\rm Auger} \rangle \sim \langle S_{\rm sim} \rangle$
$\sim \langle Q_{\rm AGN} \rangle \sim \langle Q_{\rm VCV} \rangle$
is indeed consistent with such expectation.
This means, however, that the statistics of $S$ reflect mainly on the
distribution of reference objects, rather than the deflection angle.

To further compare the Auger data with our simulations,
we plot the cumulative fraction of events, $F(\le \log S)$, versus $\log S$
for the simulated events (lines) and the Auger events (open circles)
in Figure 6.
The solid and dotted lines are for the cases of $\gamma = 2.7$ and 2.4,
respectively, and the difference between the two cases is again small.
The Kolmogorov-Smirnov (K-S) test yields the maximum difference of
$D = 0.17$, 0.23, and 0.26 between the Auger data and
the simulation data ($\gamma = 2.7$) in Models A, B, and C, respectively;
the significance level of the null hypothesis that the two
distributions are statistically identical is $P\sim 0.37$, 0.09, and 0.04
for Models A, B, and C, respectively.
So the null hypothesis that the two distributions for our simulated
events and the Auger events are statistically identical cannot 
be rejected, especially for Model A.
This would be a justification for our models of the IGMF, sources of UHECRs,
and reference objects.  
Also we see that Model A with more sources is preferred over Models B and
C with fewer sources.
But this does not necessary mean that all the AGNs would be the actual
sources of UHECRs.
We note that the number of the Auger events used, 27, is still limited.
In addition, we consider only UHE protons in this paper.
Hence, before we argue the above statements for sure, we will need more
observational events and need to know the composition of UHECRs
(see Summary and Discussion for further discussion on composition).

\subsection{Probability of Finding True Sources}

With $\langle \theta \rangle \gg \langle S_{\rm sim} \rangle$,
there is a good chance that the AGNs found closest to the direction of
UHECRs are not the actual sources of UHECRs.
To illustrate this point, we first show the distribution of $D_S$ versus
$D_{\theta}$ in Figure 7.
For some events, the closest AGNs are the actual sources.
They are represented by the diagonal line.
Around the diagonal line, a noticeable fraction of events are found.
Those are the events for which the closest AGNs are found around the
true sources; both sources and close-by AGNs are clustered as a part
of the LSS of the universe.
For the events away from the diagonal line, it is more likely that
$D_S > D_{\theta}$.
It is because there are more AGNs with larger $D_S$; away from true sources,
observed events are more likely to pick up closest AGNs with larger $D_S$.

To quantify the consequence of
$\langle \theta \rangle \gg \langle S_{\rm sim} \rangle$,
we calculated the fraction of true identification, $f$, as the ratio of
the number of events for which nearest AGNs are their true sources
to the total number of simulated events.
This is a measure of the probability to find the true sources of UHECRs,
when nearest candidates are blindly chosen (which is the best we can
do with observed data). 
In Figure 8, we show the fraction as a function of separation angle, $S$.
The fraction is largest in Model A with largest $N_{\rm src}$, 
and smallest in Model C with smallest $N_{\rm src}$.
At $S \sim 2^{\circ}$ the fraction is about 50 \% for Model A, close to
40 \% for Model B, and a little above 30 \% for Model C, but only
$20 - 30$ \% at $S = 3^{\circ}-4^{\circ}$.
As the separation angle increases, the fraction decreases gradually to
$\sim 10$ \%, indicating lower probabilities to find true sources at
larger separation angles. 
On average, we should expect that in less than 1 out of 3 events, the
true sources of UHECRs can be identified, if our model for the IGMF is
valid and UHECRs are protons.

\section{Summary and Discussion}

In the search for the nature and origin of UHECRs, understanding the
propagation of charged particles through the magnetized LSS of the
universe is important.
At present, the details of the IGMF are still uncertain, mainly due to
limited available information from observation. 
Here, we adopted a realistic model universe that was described by
simulations of cosmological structure formation;
our simulated universe represents the LSS, which is dominated by the
cosmic web of filaments interconnecting clusters and groups.
The distribution of the IGMF in the LSS of the universe was obtained with
a physically motivated model based on turbulence dynamo \citep{rkcd08}.

To investigate the effects of the IGMF on the arrival direction of
UHECRs, we further adopted the following models.
Virtual observers of about 1400 were placed at groups of galaxies, which
represent statistically the Local Group in the simulated model universe.
Then, we set up a set of about 500 AGN-like ``reference objects'' within
75 Mpc from each observer, at clusters of galaxies (deep gravitational
potential wells) along the LSS.
They represent a class of astronomical objects with which we performed
a correlation analysis for simulated UHECR events.
We considered three models, in which subsets of the reference objects were 
selected as AGN-like sources of UHECRs (see Table 1).
UHE protons of $E \ge 60$ EeV with power-low energy spectrum were
injected at those sources, and the trajectories of UHE protons in the
magnetized cosmic web were followed. 
At observer locations, the events with $E \ge 60$ EeV from sources
within a sphere of radius 75 Mpc were recorded and analyzed.

To characterize the clustering of the reference objects, we calculated
the angular distance, $Q$, from a given reference object to its nearest
neighbor.
The mean value for our model AGNs in the simulated universe is
$\langle Q_{\rm AGN} \rangle = 3.68^{\circ} \pm 1.66^{\circ}$,
while that for 442 AGNs from the VCV catalog is 
$\langle Q_{\rm VCV} \rangle = 3.55^{\circ}$.
This demonstrates that the two samples have a similar degree of clustering
and are highly structured (e.g. $\langle Q_{\rm iso}\rangle \approx 11^{\circ}$
for the isotropic distribution). 

With our model IGMF, the deflection angle, $\theta$, between the arrival
direction of UHE protons and the sky position of their actual sources, is
quite large with the mean $\langle \theta \rangle \sim 14 - 17.5^{\circ}$
and the median $\tilde \theta \sim 7 - 10^{\circ}$, depending on models
with different numbers of sources (see Table 1).
On the other hand, the separation angle between the arrival direction
and the sky position of nearest reference objects is substantially smaller
with the mean $\langle S_{\rm sim} \rangle \sim 3.5 - 4^{\circ}$ and
the median $\tilde S_{\rm sim} = 2.8 - 3.5^{\circ}$.
That is, we found that while
$\langle \theta \rangle \sim 4 \langle S_{\rm sim} \rangle$,
$\langle S_{\rm sim} \rangle$ is similar to  $\langle Q_{\rm AGN} \rangle$.
For the Auger events of highest energies in \citet{auger08a}, with 442
nearby AGNs from the VCV catalog as the reference objects, the mean
separation angle is $\langle S_{\rm Auger} \rangle = 3.23^{\circ}$ for
the 26 events,  excluding one event with large $S$ ($ \approx 27^{\circ}$),
while $\langle S_{\rm Auger} \rangle =4.13^{\circ}$ for all the 27 events. 
Hence, $\langle S_{\rm Auger} \rangle \sim \langle Q_{\rm VCV} \rangle$
$\sim \langle S_{\rm sim} \rangle \sim \langle Q_{\rm AGN} \rangle $.  
This implies that the separation angle from the Auger data would be
determined primarily by the distribution of reference objects (AGNs),
and may not represent the true deflection angle.

We further tested whether the distributions of separation angle, $S$, for
our simulated events and for the Auger events are statistically comparable
to each other.
According the Kolmogorov-Smirnov test for the cumulative fraction of
events, $F(\le \log S)$, versus $\log S$, the significance level of the
null hypothesis that the two distributions are drawn from the identical
population is as large as $P\sim 0.37$ for Model A (see Table 1).
Thus, we argued that our simulation data, especially in Model A, are in
a fair agreement with the Auger data.
This test also showed that the model with more sources (Model A) is
preferred over the models with fewer sources (Models B and C).

The fact that $\langle \theta \rangle \gg \langle S_{\rm sim} \rangle$
implies that the AGNs found closest to the direction of UHECRs may not be
the true sources of UHECRs.
We estimated the probability of finding the true sources of UHECRs,
when nearest reference objects are blindly chosen:
$f(S)$ is the ratio of the number of
true source identifications to the total number of simulated events.
This probability is $\sim 50 - 30$ \% at $S \sim 2^{\circ}$, but
decreases to $\sim 10$ \% at larger separation angle.
On average, in less than 1 out of 3 events, the true sources of UHECRs
can be identified in our simulations, when nearest reference objects
are chosen.

The distribution of $\theta$ versus $D_{\theta}$ shows a bimodal pattern
in which  $\theta$ is on average larger either for nearby sources
(for $D_{\theta} \la 15$ Mpc) or for distant sources (for
$D_{\theta} \ga 30$ Mpc) with the minimum at the intermediate distance
of $D_{\theta,{\rm min}} \sim 20 - 30$ Mpc.
The distribution of $S$ versus $D_s$ shows a similar, but weaker sign of 
the bimodal pattern.
This behavior is a characteristic signature of the magnetized cosmic web
of the universe, where filaments are the most dominant structure.
When a large number of super-GZK events are accumulated, we may find 
the signature of the cosmic web of filaments in the $S$ versus $D_s$
distribution.   

Finally, we address the limitations of our work. 
(1) We worked in a simulated universe with specific models for the elements
such as the IGMF, observers, sources, and reference objects, but not in
the real universe.
So we could make only statistical statements.
(2) It has been shown previously that adopting different models for the
IGMF, very different deflection angles are obtained
\citep[see][]{sme03,dgst05,dkrc08}.
We argue that our model for the IGMF is most plausible, since it is a
physically motivated model based on turbulence dynamo without involving
an arbitrary normalization \citep{rkcd08}.
Nevertheless, our IGMF model should be confirmed further by observation.
(3) The sources of UHECRs may not be objects like AGNs, but could be
objects extinguished a while ago, such as gamma-ray bursts
\citep[see, e.g.,][]{viet95,waxm95}, or sources spread over space like
cosmological shocks \citep[see, e.g.,][]{krj96,krb97}.
The injection energy spectrum of power-law with cut-off at an
arbitrary maximum energy (see Section 2.6) would be unrealistic.
The IGMF in the Local Group (see Paper I), although currently little is
known,  might be strong enough to substantially deflect the trajectories
of UHECRs.
All of those will have effects on the quantitative results,
which should be investigated further.
(4) Recently, the Auger collaboration disclosed the analysis, which
suggests a substantial fraction of highest energy UHECRs might be iron
nuclei \citep{auger07b,auger09d}.
This is in contradiction with the analysis of the HiRes data, which
indicates highest energy UHECRs would be mostly protons \citep{st07}.
The issue of composition still needs to be settled down among experiments.
Iron nuclei, on the way from sources to us, suffer much larger deflection
than protons.
Hence, if a substantial fraction of UHECRs is iron, some of our findings
will change, a question which should be investigated in the future.

\acknowledgements
The authors would like to thank P. L. Biermann for stimulating discussion.
The work was supported by the Korea Research Foundation (KRF-2007-341-C00020)
and the Korea Foundation for International Cooperation of Science and
Technology (K20702020016-07E0200-01610).

\clearpage

\begin{figure}
\epsscale{0.8}
\vskip 0cm
\plotone{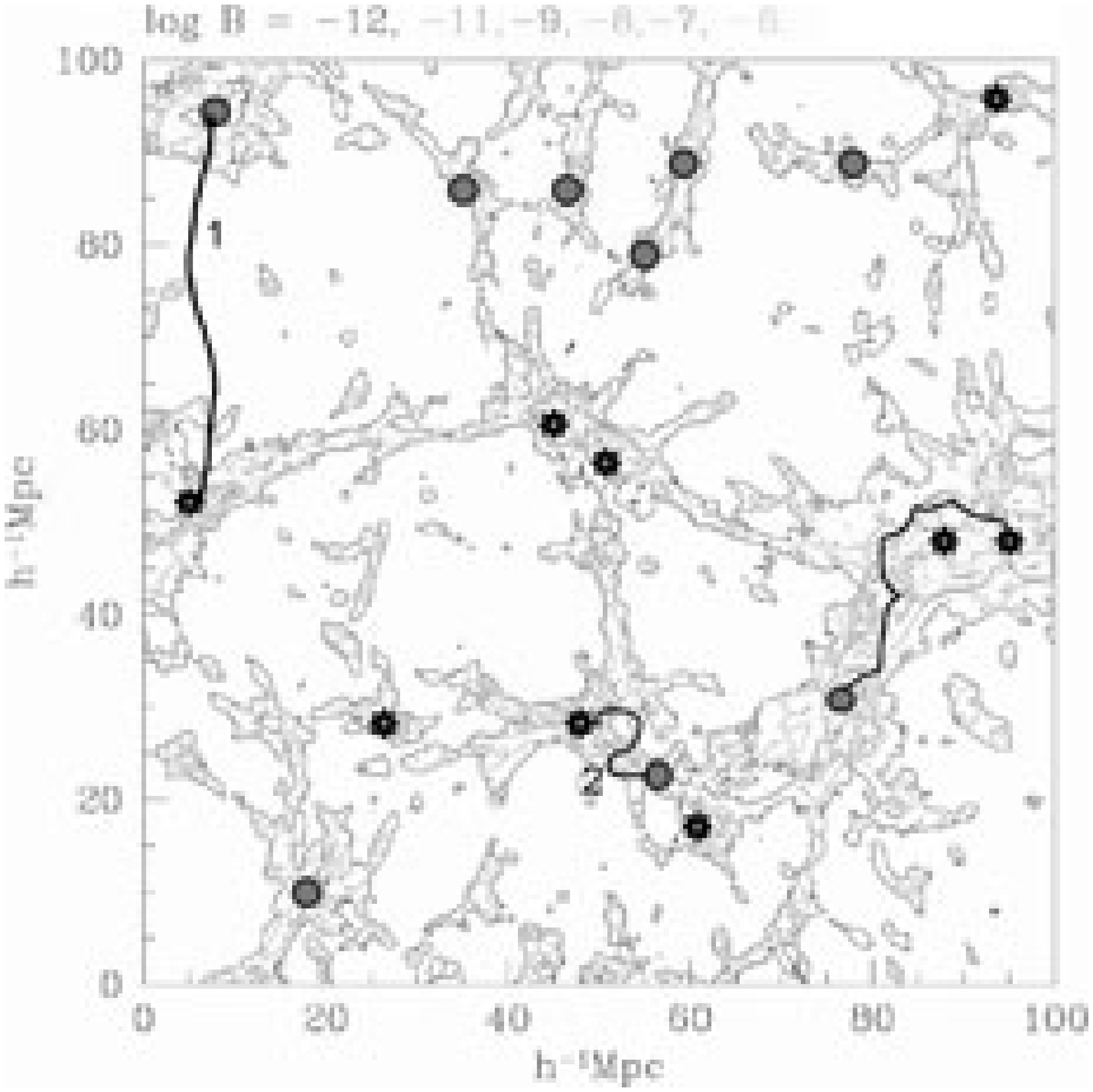}
\vskip -0.2cm
\caption{Distribution of the IGMF in a two-dimensional slice of
(143 Mpc)$^2$ in the simulated universe.
Locations of virtual observers (circles) and model AGNs (stars) are
schematically marked at clusters and groups of galaxies.
Paths of UHECRs from sources to observers are also schematically drawn.}
\end{figure}

\clearpage

\begin{figure}
\epsscale{0.8}
\vskip 0cm
\plotone{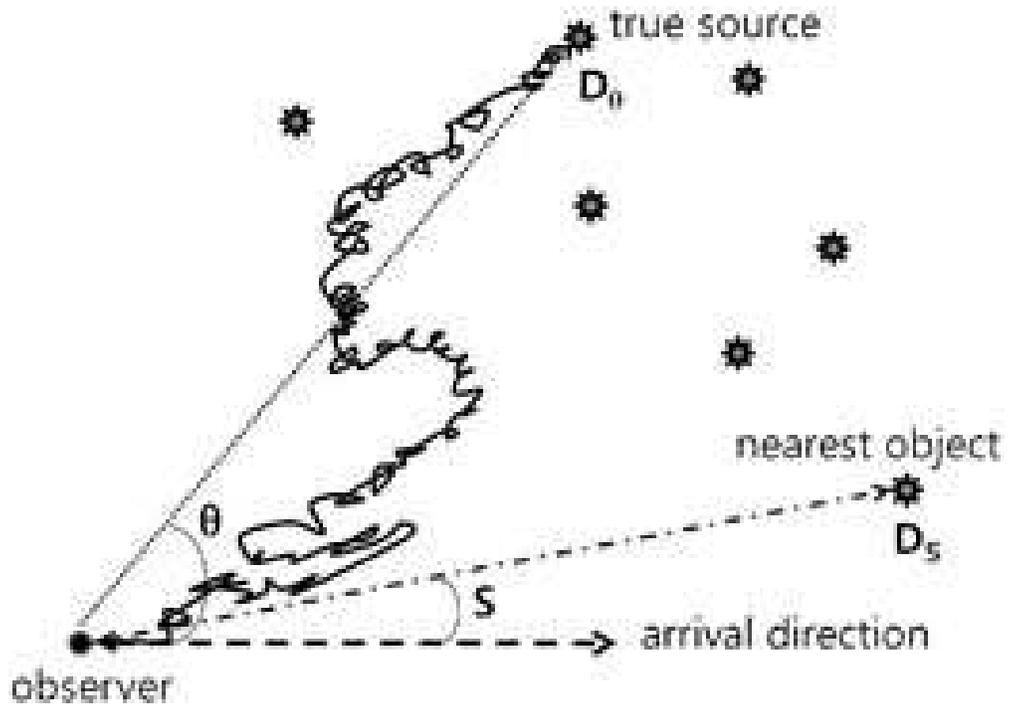}
\vskip -0.0cm
\caption{Geometry of the deflection angle, $\theta$, the separation angle,
$S$, the distance to the true source, $D_{\theta}$, and the distance to the
nearest object, $D_S$.
The path of the UHECR event from the source to observer is schematically
drawn.}
\end{figure}

\clearpage

\begin{figure}
\epsscale{0.46}
\vskip -0.8cm
\plotone{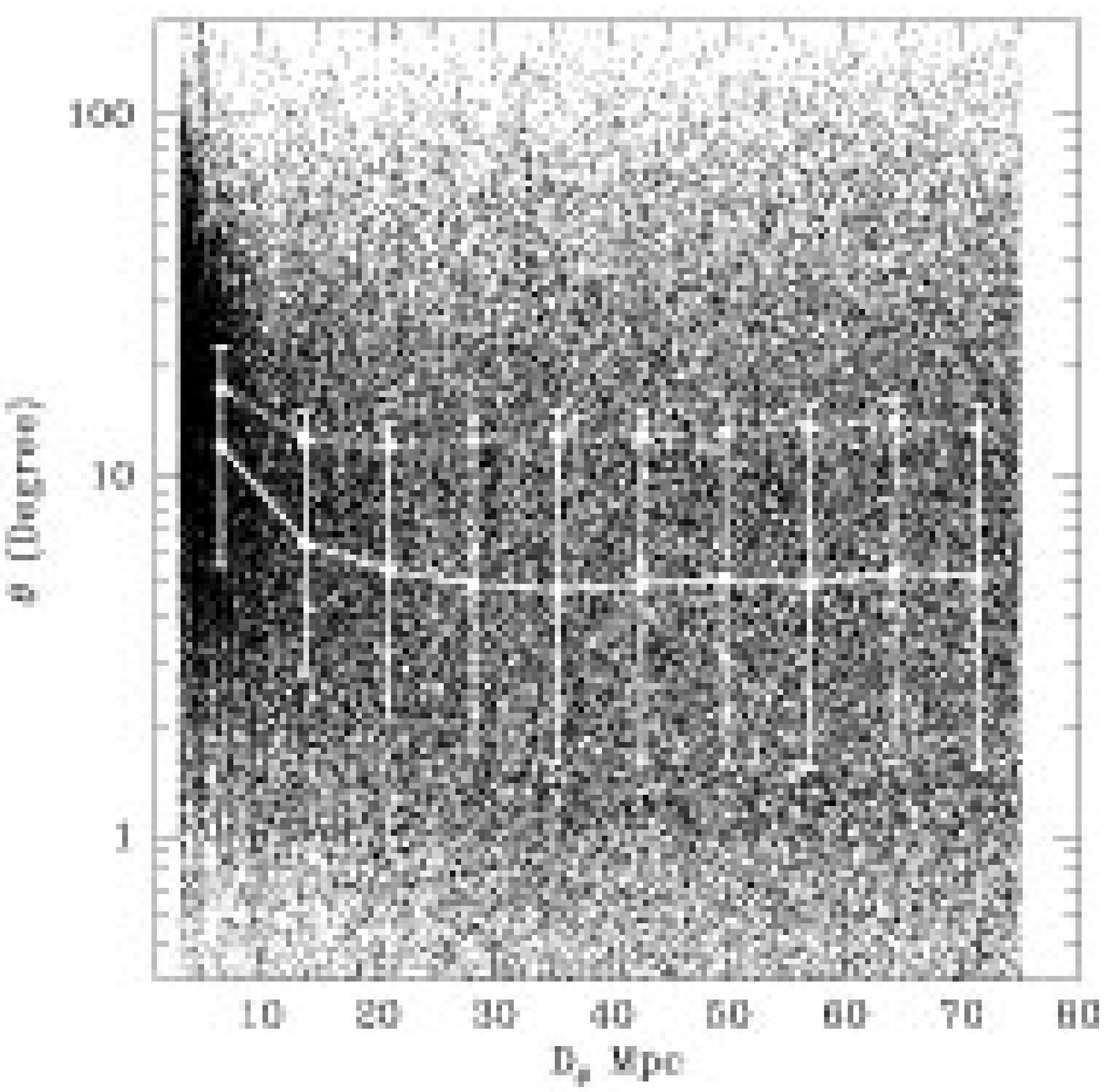}
\vskip -0.6cm
\plotone{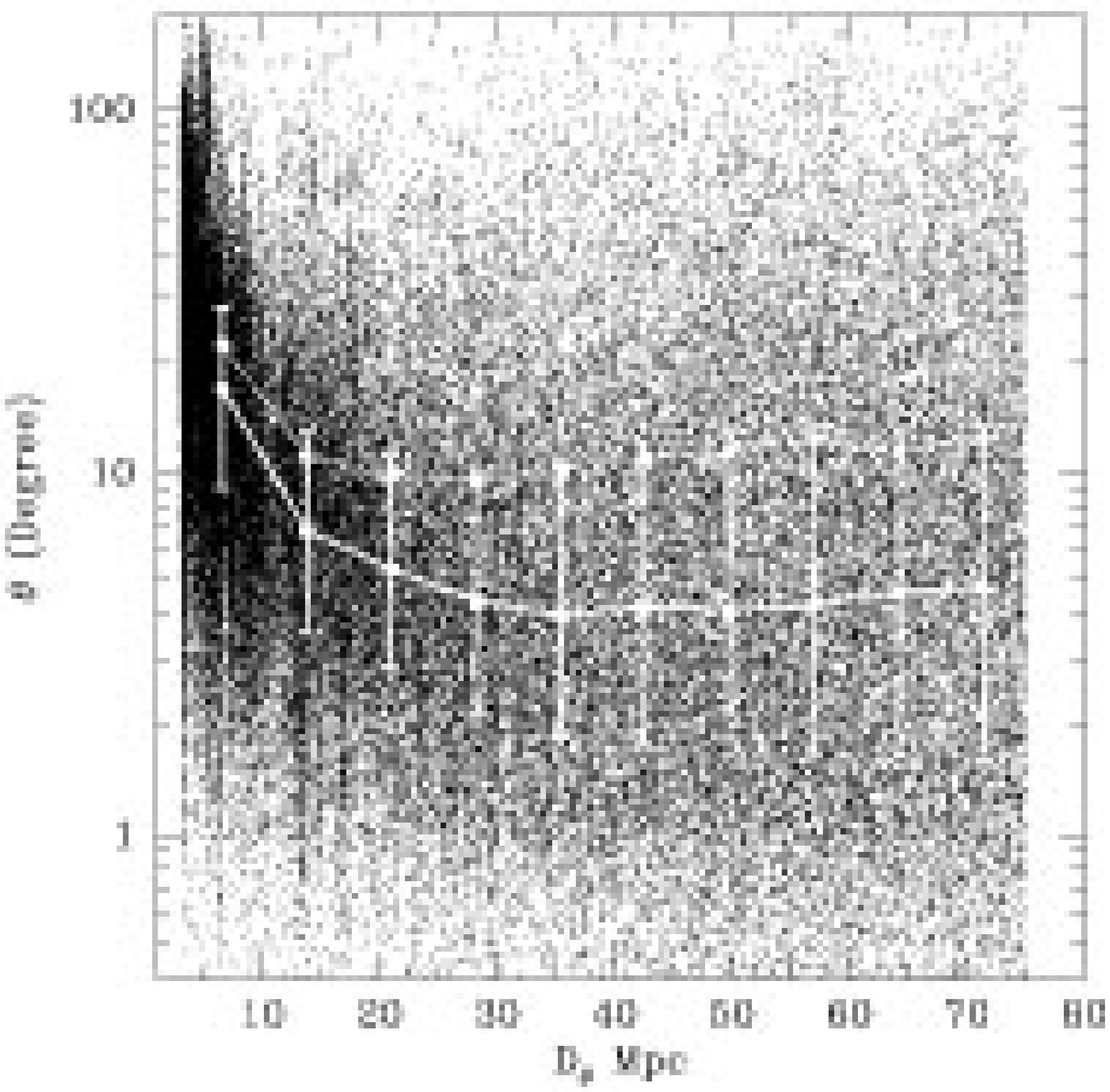}
\vskip -0.6cm
\plotone{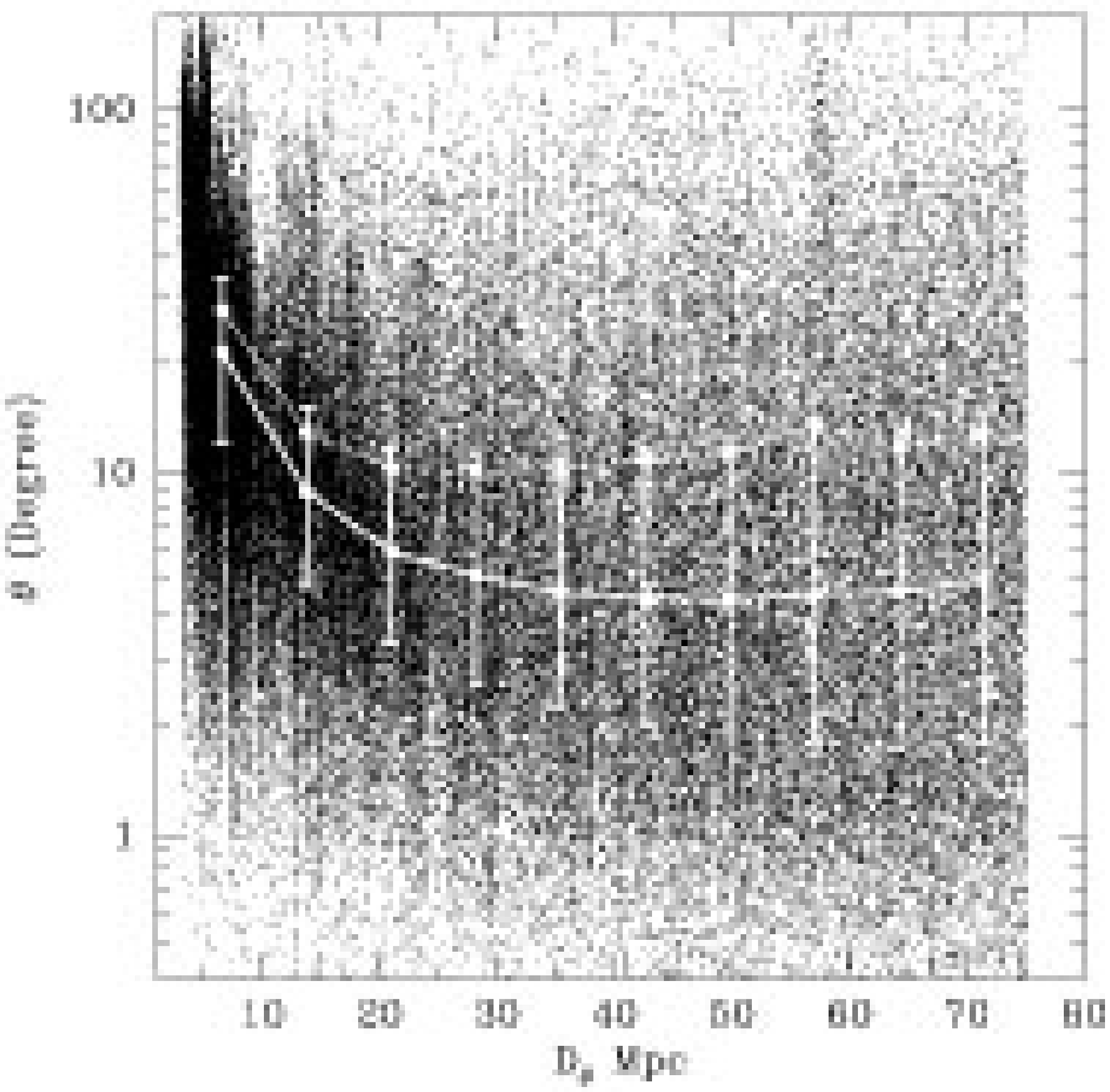}
\vskip -0.6cm
\caption{Distribution of deflection angles $(\theta)$ as a function of
distance to true sources ($D_{\theta}$).
Dots represent UHE proton events recorded in our simulations.
Circles connected with dotted lines (upper) and solid lines (lower) show
the mean and median values, respectively.
Vertical lines connect the marks of first and third quartiles in given
$D_{\theta}$ bins.
Top, middle, and bottom panels are for Model A, Model B, and Model C,
respectively.
The cases of $\gamma = 2.7$ are shown.}
\end{figure}

\clearpage

\begin{figure}
\epsscale{0.48}
\vskip -0.8cm
\plotone{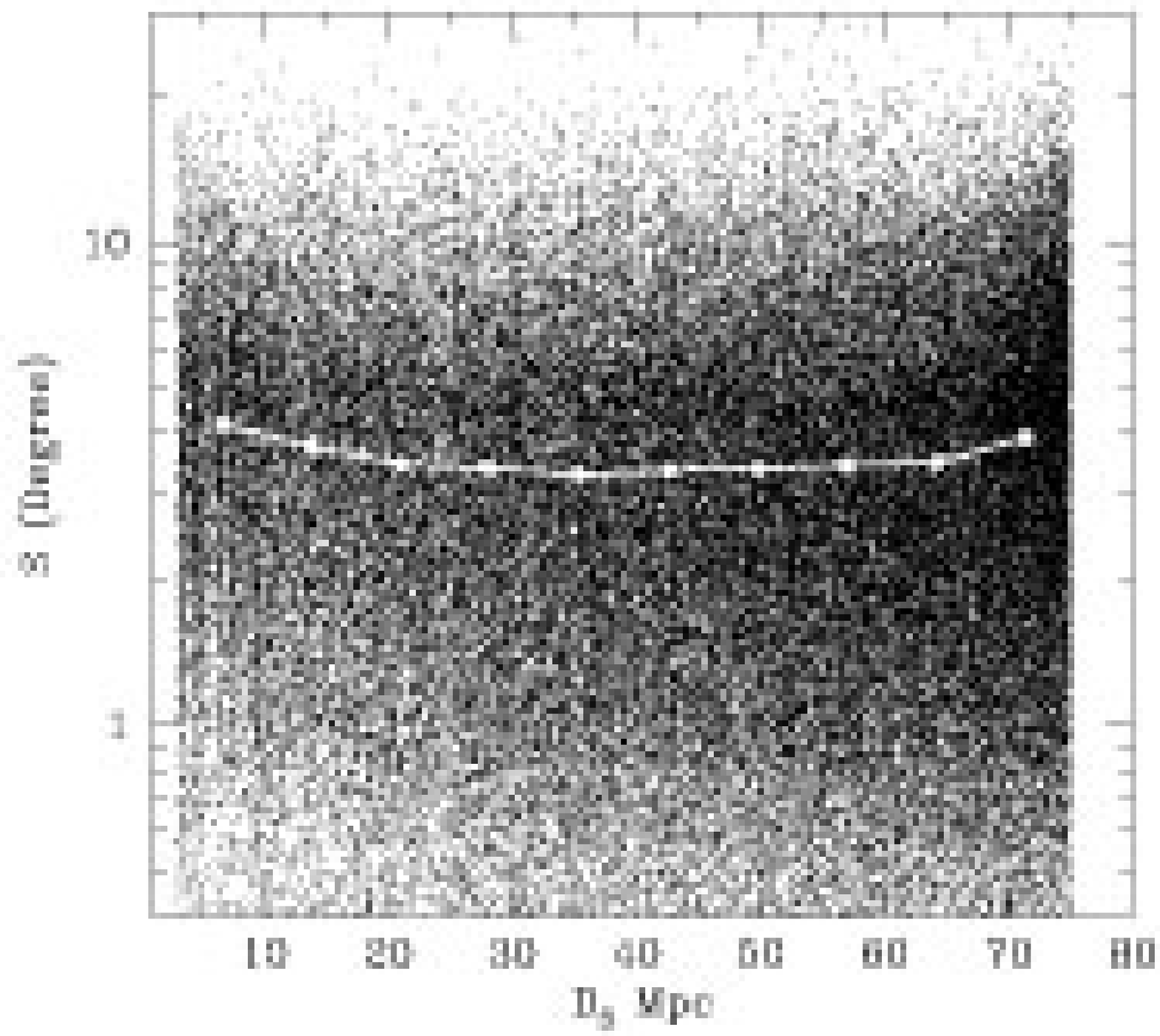}
\vskip -0.7cm
\plotone{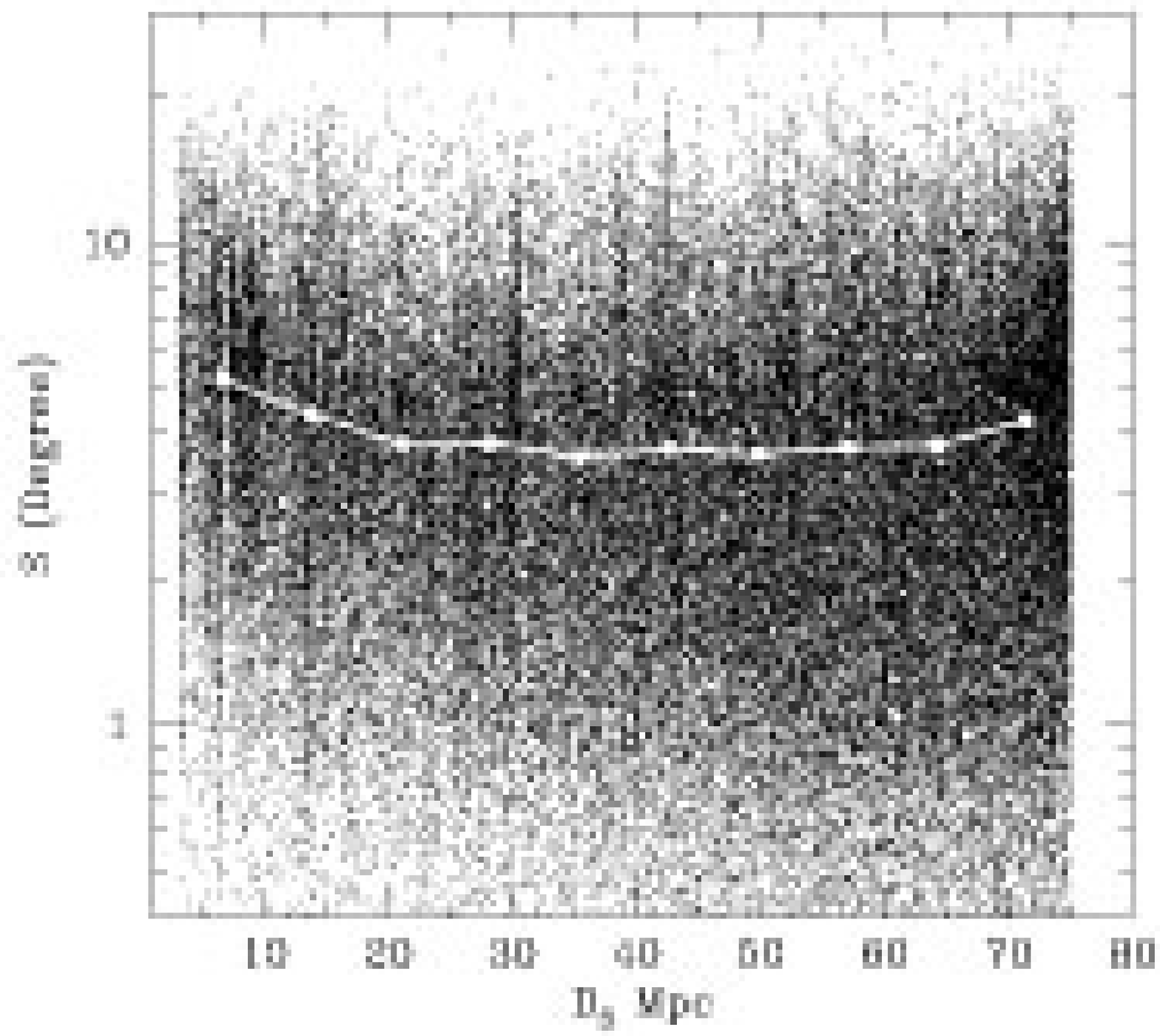}
\vskip -0.7cm
\plotone{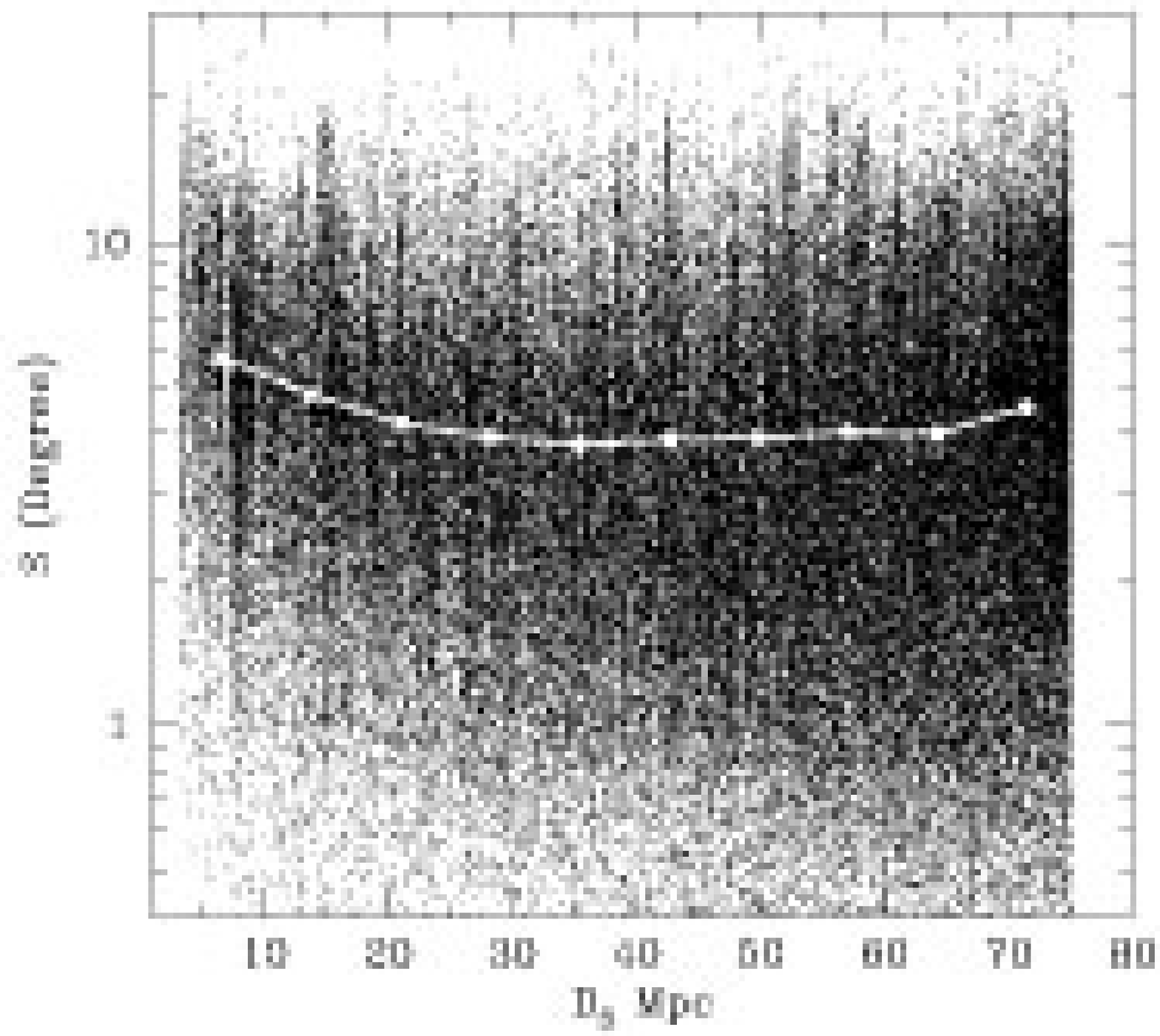}
\vskip -0.7cm
\caption{Distribution of separation angles between the directions of
UHE protons and nearest model AGNs ($S$) as a function of distance to
nearest model AGNs ($D_S$).
Dots represent UHE proton events recorded in our simulations.
Circles connected with solid lines show the mean values.
Top, middle, and bottom panels are for Model A, Model B, and Model C,
respectively.
The cases of injection spectral index $\gamma = 2.7$ are shown.}
\end{figure}

\clearpage

\begin{figure}
\epsscale{0.44}
\vskip -1.cm
\plotone{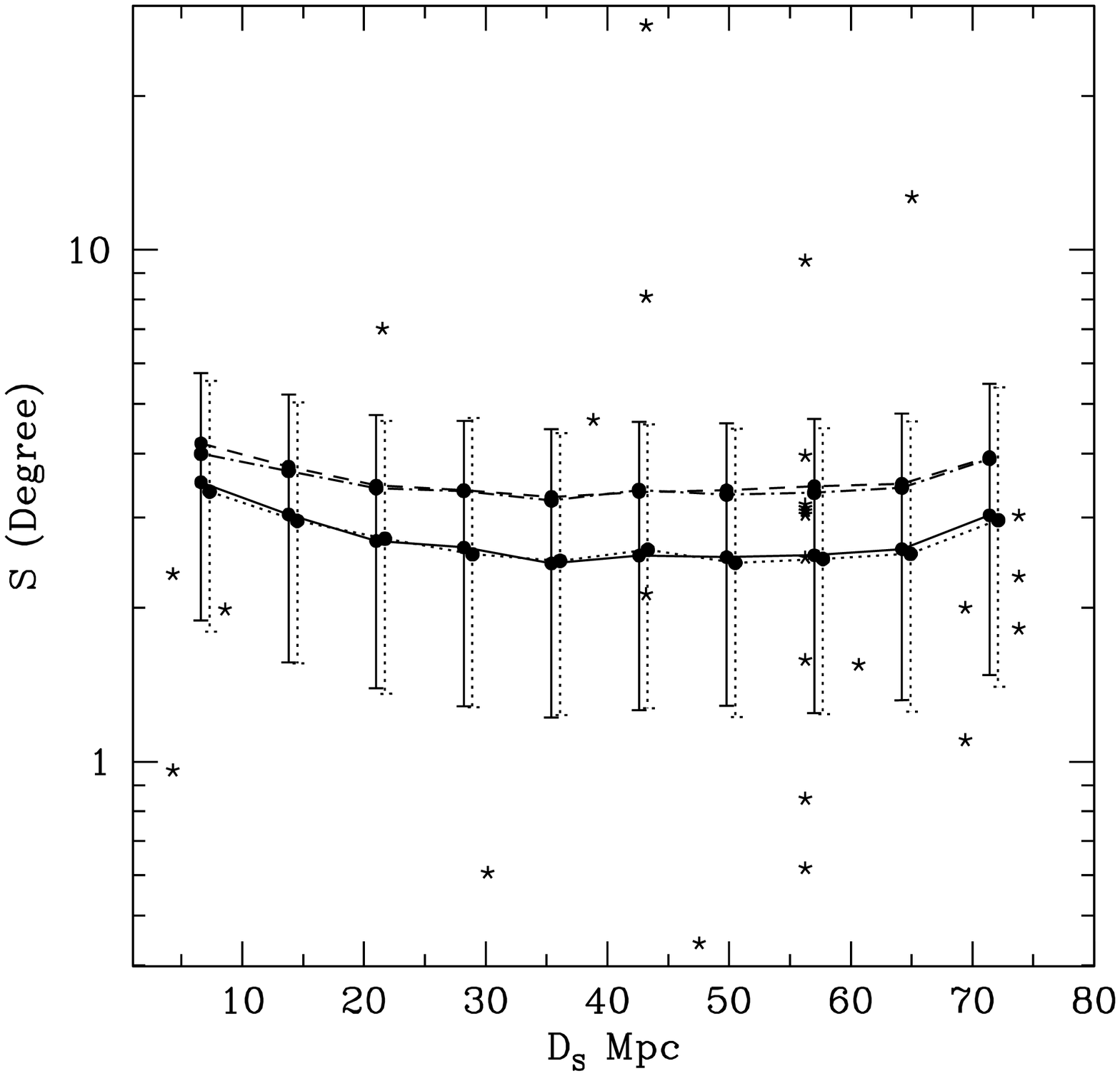}
\vskip -0.6cm
\plotone{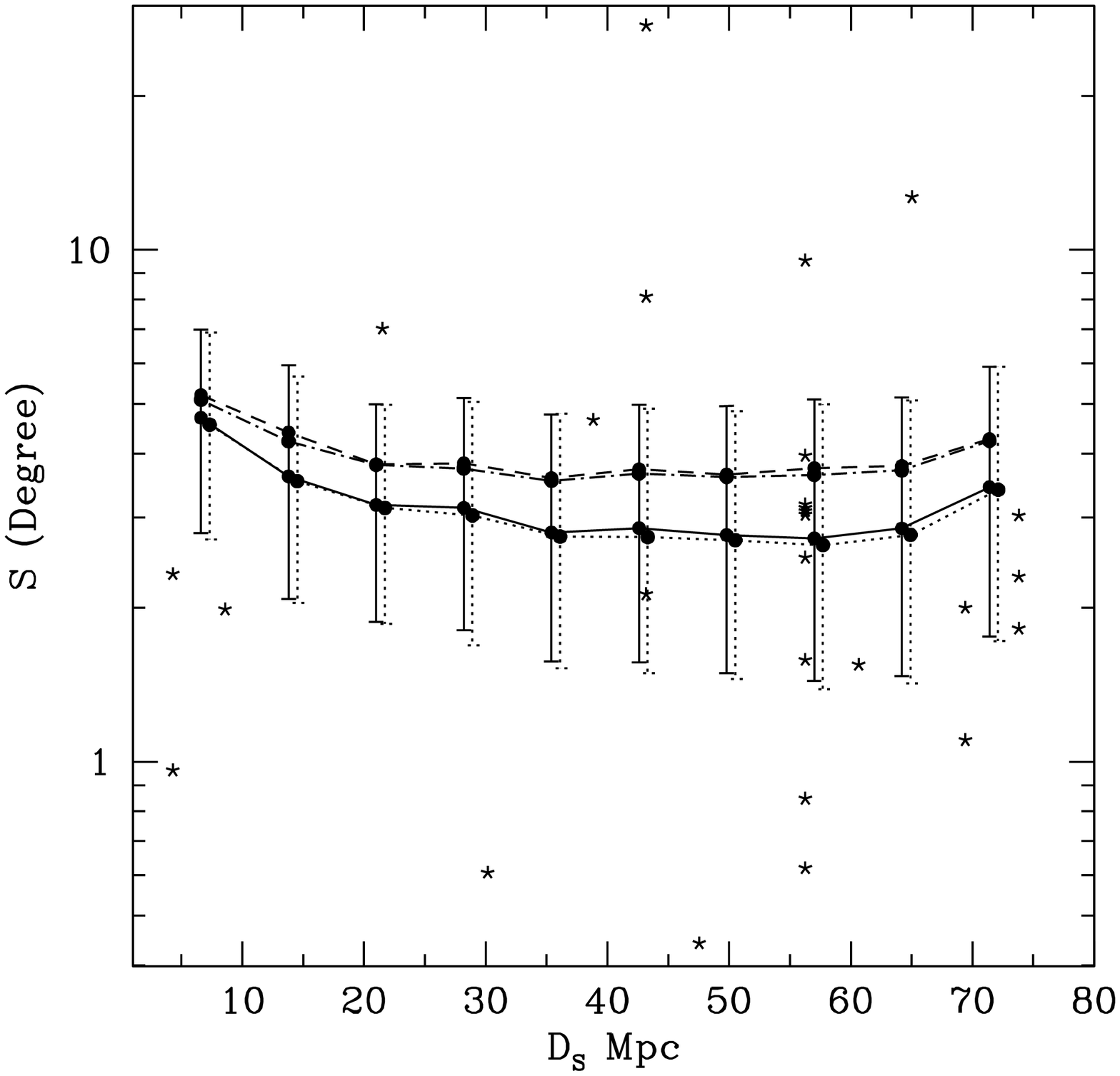}
\vskip -0.6cm
\plotone{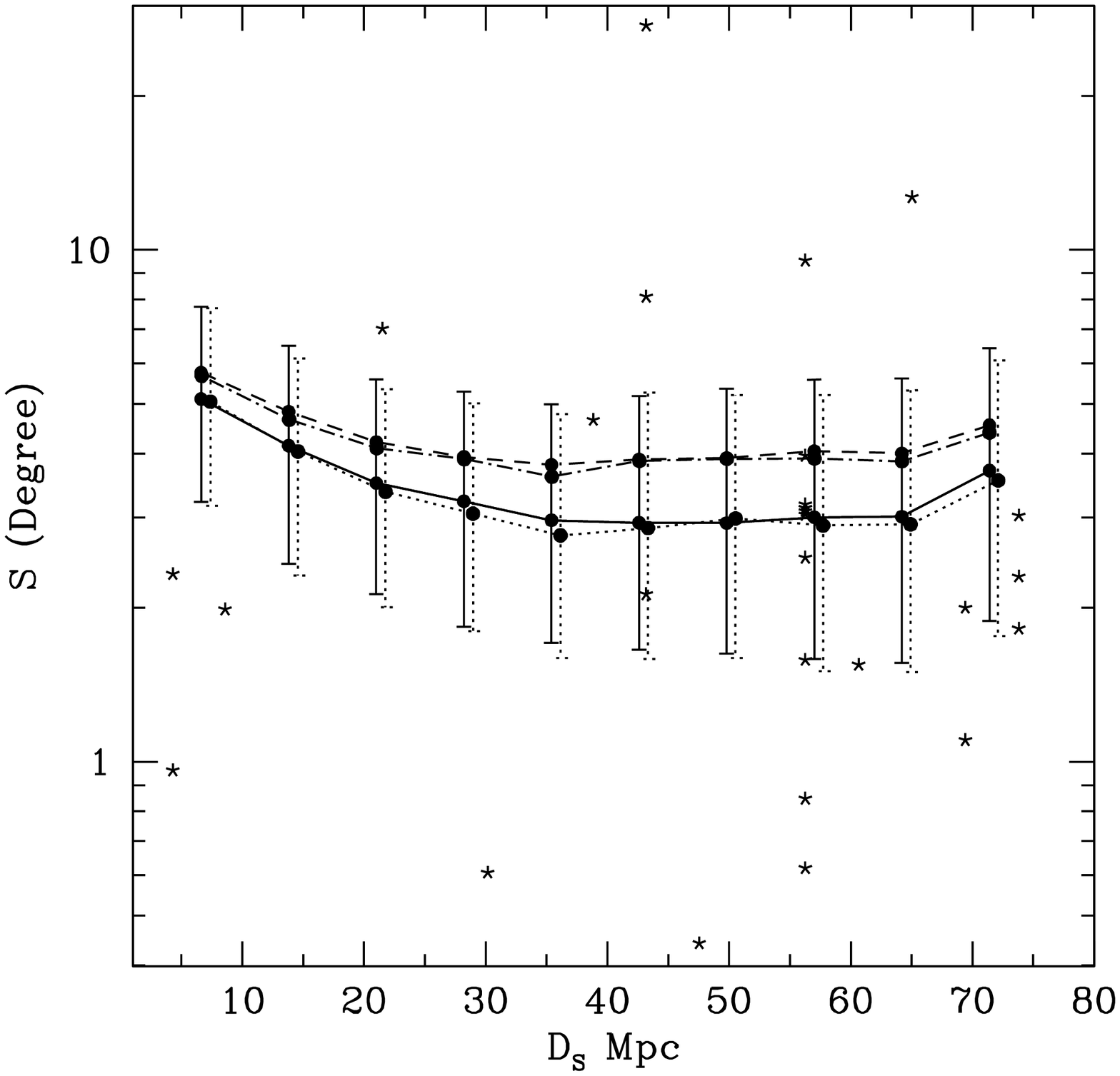}
\vskip -0.7cm
\caption{Mean (upper circles) and median (lower circles) values of $S$ as
a function of $D_S$ for UHE proton events in our simulations.
Vertical lines connect the marks of first and third quartiles in given
$D_S$ bins.
Circles connected with dashed lines (upper) and solid lines (lower) are
for $\gamma = 2.7$, and those connected with dot-dashed lines (upper) and
dotted lines (lower) are for $\gamma = 2.4$.
The mean $S$ for $\gamma = 2.7$ are the same as those in Figure 2.
The median and quartiles for $\gamma = 2.4$ are horizontally shifted for
better visibility.
Asterisks denote $S$ for the 27 Auger events of highest energies, published
in \citet{auger08a}, with nearby AGNs from the VCV catalog.
Top, middle, and bottom panels are for Model A, Model B, and Model C,
respectively.}
\end{figure}

\clearpage

\begin{figure}
\epsscale{0.5}
\vskip -1.3cm
\plotone{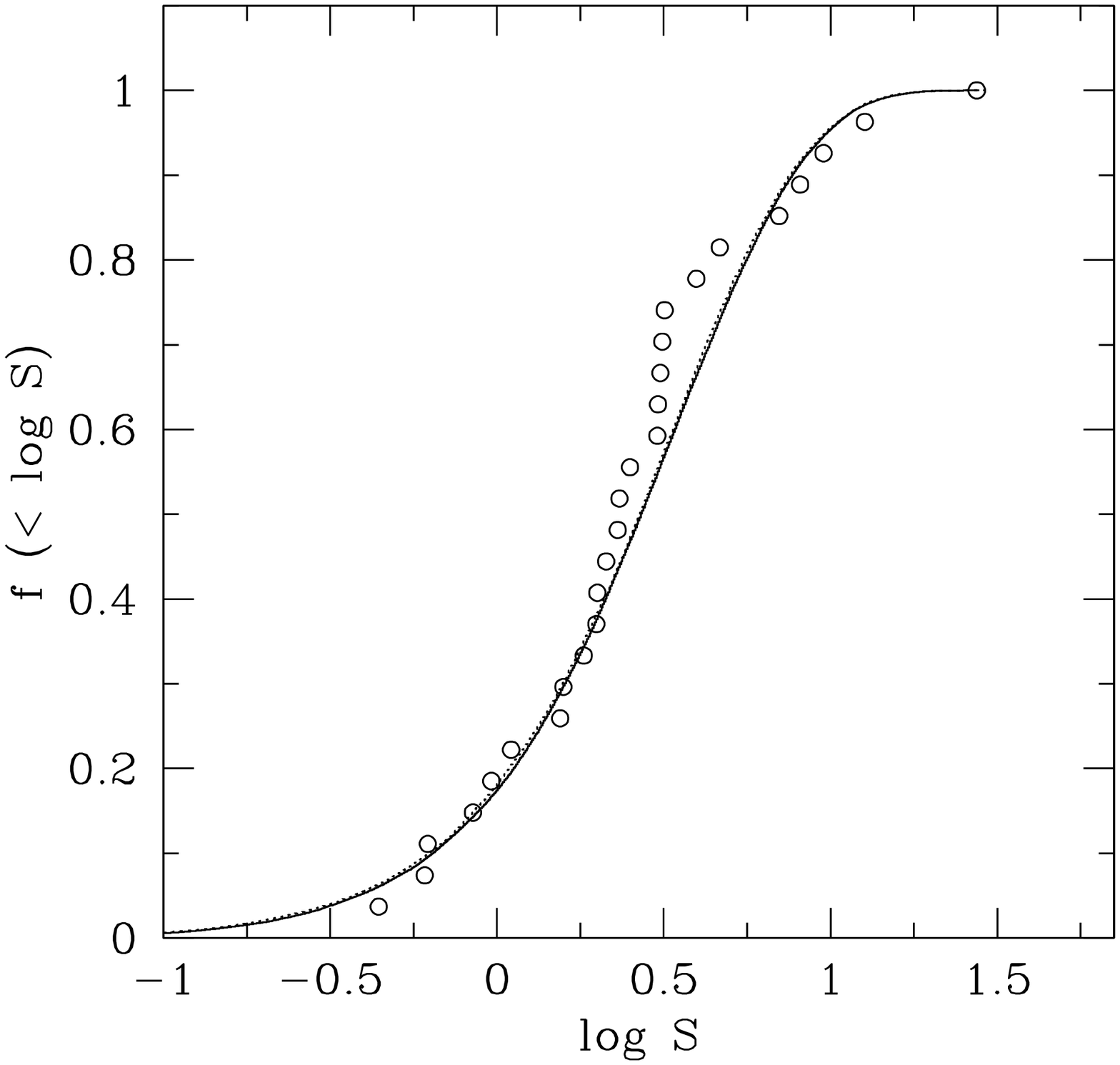}
\vskip -0.7cm
\plotone{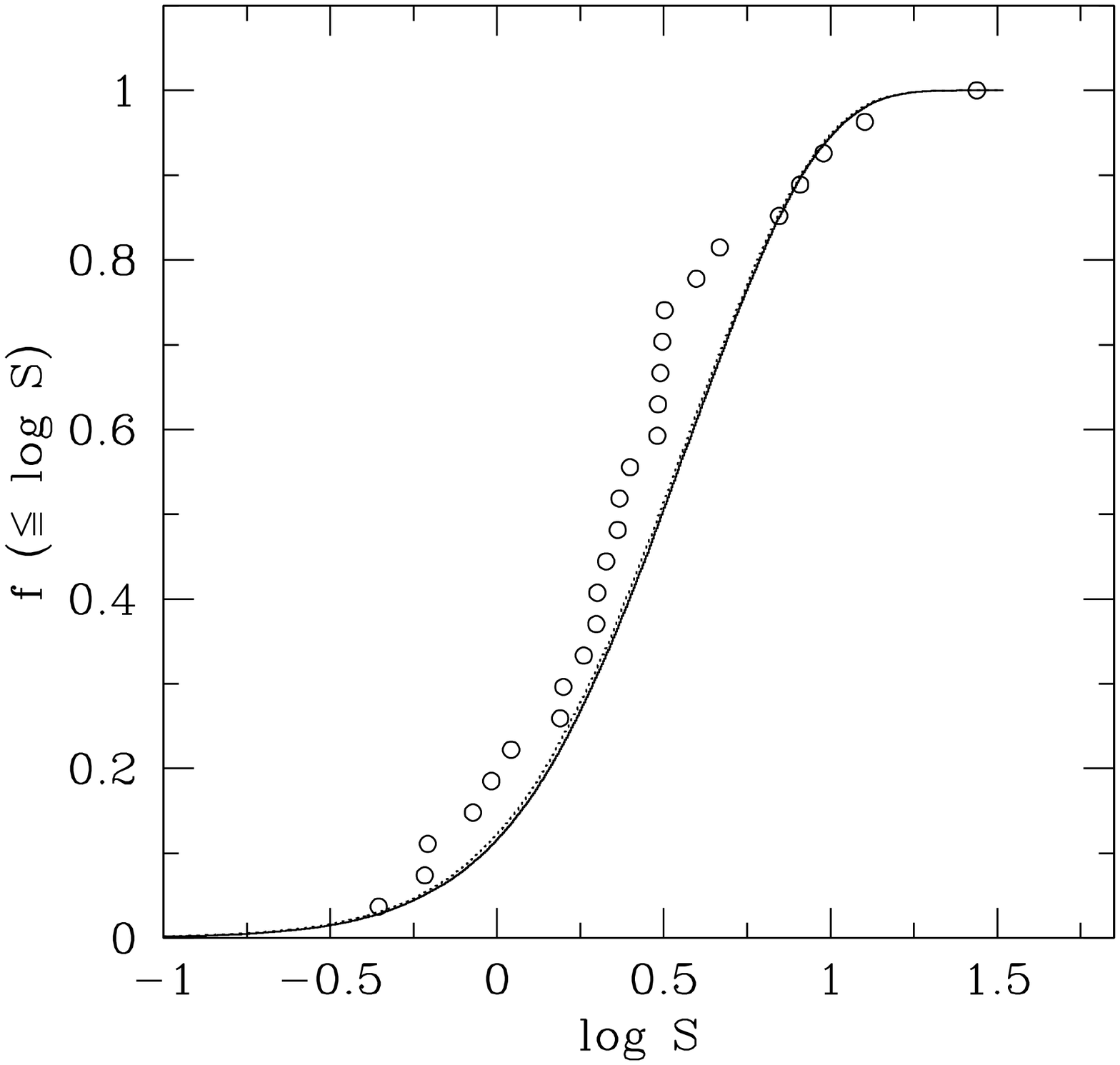}
\vskip -0.7cm
\plotone{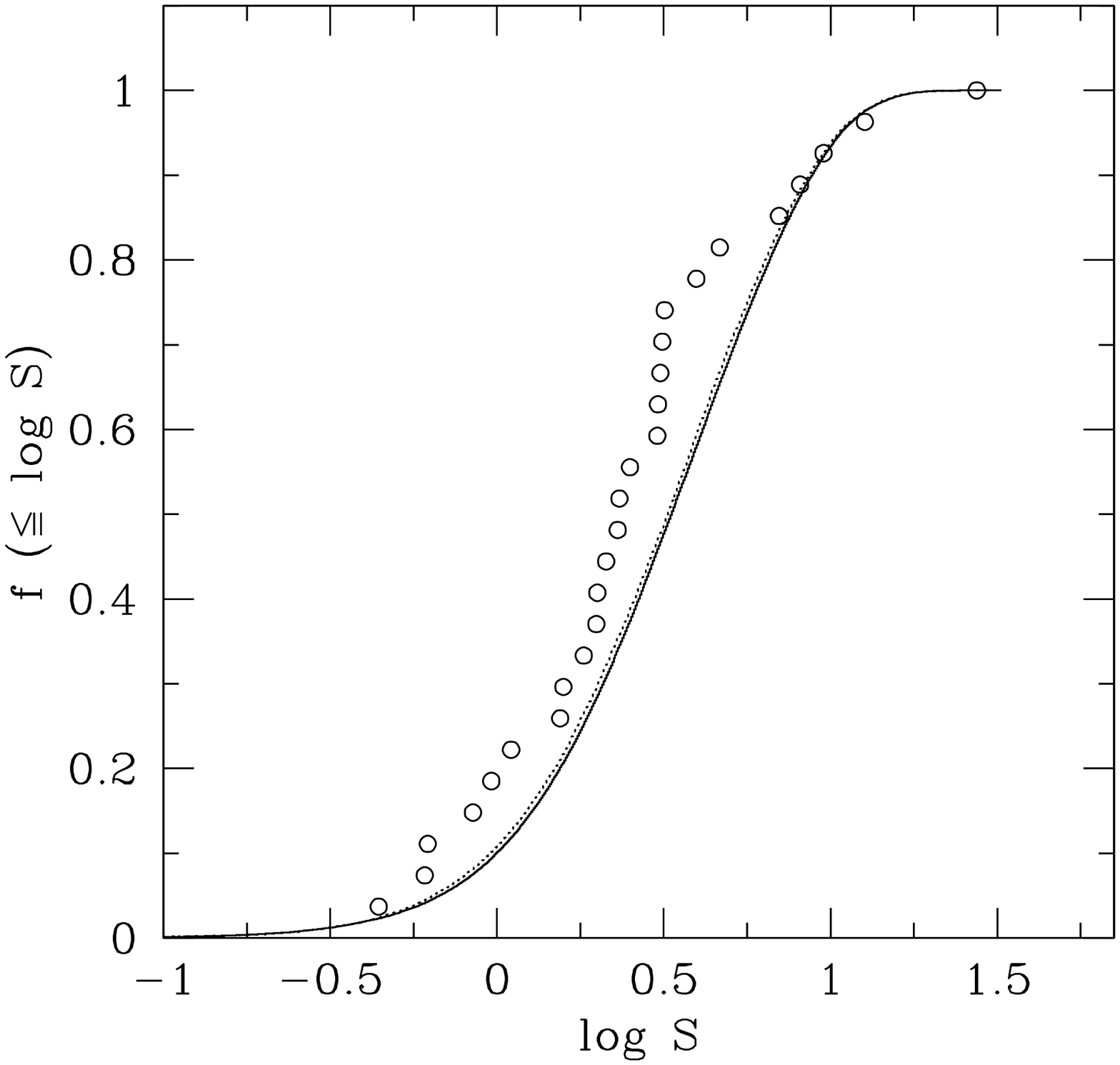}
\vskip -0.7cm
\caption{Cumulative fraction of UHECR events with separation angle smaller
than $S$.
Solid and dotted lines denote the results calculated with the UHE proton
events in our simulations for $\gamma = 2.7$ and $2.4$, respectively.
Top, middle, and bottom panels are for Model A, Model B, and Model C,
respectively.
Open circles denote the result for the 27 Auger events \citep{auger08a}
with nearby AGNs from the VCV catalog.}
\end{figure}

\clearpage

\begin{figure}
\epsscale{0.6}
\vskip 0cm
\plotone{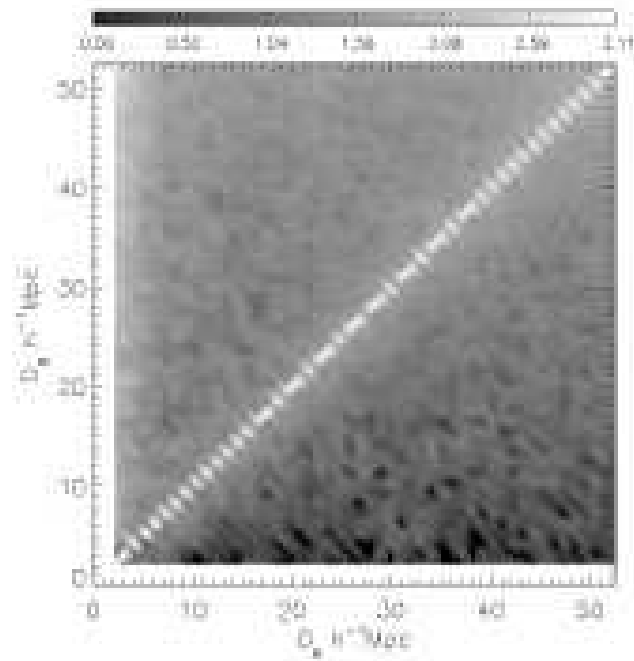}
\vskip -0.7cm
\caption{Distance to nearest model AGNs, $D_S$, versus distance to true
sources, $D_{\theta}$, for the UHECR events in Model A.
The case of $\gamma = 2.7$ is shown.
Color codes the number of events in $\log_{10}$ scale in bins of
$\Delta D_\theta \times \Delta D_S$.
The maximum distance is 75 Mpc for both $D_S$ and $D_{\theta}$.}
\end{figure}

\clearpage

\begin{figure}
\epsscale{0.55}
\vskip 0cm
\plotone{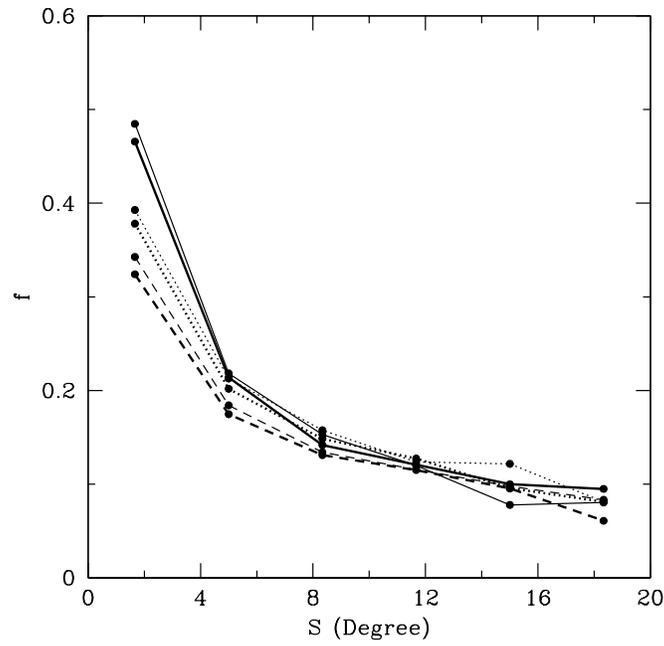}
\vskip -0.2cm
\caption{Fraction ($f$) of UHE proton events recorded in our simulations,
for which their true sources are identified as closest AGNs in the sky,
as a function of $S$.
Dashed, dotted, and solid lines are for Model A, Model B, and Model C,
respectively.
Heavy and light lines denote the fractions for $\gamma = 2.7$ and $2.4$,
respectively.}
\end{figure}

\end{document}